\documentclass[apj]{emulateapj}
\usepackage{hyperref}
\usepackage{amsmath}
\usepackage{color}

\shorttitle{AGN in Galaxy Clusters at $z\sim1$}
\shortauthors{W. Mo et al.}

\begin{document}

\title{The Massive and Distant Clusters of WISE Survey IV: The Distribution of Active Galactic Nuclei in Galaxy Clusters at $z\sim1$ }

\author{
Wenli Mo\altaffilmark{1},
Anthony Gonzalez\altaffilmark{1},
Daniel Stern\altaffilmark{2},
Mark Brodwin\altaffilmark{3},
Bandon Decker\altaffilmark{3},
Peter Eisenhardt\altaffilmark{2},
Emily Moravec\altaffilmark{1},
S. A. Stanford\altaffilmark{4},
and
Dominika Wylezalek\altaffilmark{5,6}
}

\altaffiltext{1}{Department of Astronomy, University of Florida, Bryant Space Science Center, Gainesville, FL 32611}
\altaffiltext{2}{Jet Propulsion Laboratory, California Institute of Technology, Pasadena, CA 91109}
\altaffiltext{3}{Department of Physics and Astronomy, University of Missouri, 5110 Rockhill Road, Kansas City, MO 64110}
\altaffiltext{4}{Department of Physics, University of California, Davis, One Shields Avenue, Davis, CA 95616, USA}
\altaffiltext{5}{Department of Physics and Astronomy, Johns Hopkins University, Bloomberg Center, 3400 N. Charles St, Baltimore, MD 21218, USA}
\altaffiltext{6}{European Southern Observatory, Karl-Schwarzschildstrasse 2, 85748 Garching, Germany}

\begin{abstract}

We present an analysis of the radial distribution of Active Galactic Nuclei (AGN) in $2300$ galaxy clusters from the Massive and Distant Clusters of {\it WISE} Survey (MaDCoWS). MaDCoWS provides the largest coverage of the extragalactic sky for a cluster sample at $z\sim1$. We use literature catalogs of AGN selected via optical, mid-infrared (MIR), and radio data, and by optical-to-MIR (OIR) color. Stacking the radial distribution of AGN within the $6\arcmin$ of the centers of MaDCoWS galaxy clusters, we find a distinct overdensity of AGN within $1\arcmin$ of the galaxy cluster center for AGN of all selection methods. The fraction of red galaxies that host AGN as a function of clustercentric distance is, however, dependent on the AGN selection. The fraction of red galaxies in cluster environments that host AGN selected by optical signatures or blue OIR color is at a deficit compared to the field, while MIR-selected and red OIR color AGN are enhanced in the centers of clusters when compared to field levels. The radio-selected AGN fraction is more than $2.5$ times that of the field, implying that the centers of clusters are conducive to the triggering of radio emission in AGN. We do not find a statistically significant change in the AGN fraction as a function of cluster richness. We also investigate the correlation of central radio activity with other AGN in galaxy clusters. Clusters with radio activity have more central AGN than radio-inactive clusters, implying that central cluster radio activity and AGN triggering may be linked.

\end{abstract}


\maketitle

\section{Introduction}

All galaxies are thought to undergo active phases where the central supermassive black hole (SMBH) accretes galactic material. Active Galactic Nuclei (AGN) are believed to play a role in regulating star formation and transforming blue, star forming galaxies into red, passive galaxies \citep[e.g.][]{hopkins08,hickox09}.
Evidence also points to a co-evolution of supermassive black hole mass and host galaxy properties such as bulge mass and velocity dispersion \citep[e.g.][]{magorrian98, silk98}, implying a connection between the evolution of the galaxy and AGN.

AGN fueling is dependent on both the gas supply available to a galaxy, and the efficiency with which that gas can be deposited on the central SMBH.  The former is influenced by the large-scale environment in which a galaxy resides, while the latter depends upon whether mergers \citep[e.g.,][]{hopkins08} or secular processes such as  bar instabilities and galaxy interactions \citep[e.g.,][]{goulding14} are the dominant mechanisms for triggering AGN. The dense, rich environment of galaxy clusters, with high densities of galaxies and intracluster material, provides a unique regime in which to study the impact of large-scale environmental factors. In the outer regions of galaxy clusters, the cluster environment can increase merger rates and the frequency of tidal perturbations as compared to the field, triggering AGN. Meanwhile, in the cluster cores, the intracluster medium (ICM) can quench galactic star formation and AGN fueling by removing the fuel supply. The ICM can unbind gas from the halos of infalling galaxies \citep[strangulation;][]{larson80}, while leaving the cold central gas intact, stifling the galaxy's ability to accrete new cold gas after depleting its initial cold gas content. It can also remove cooler gas within the galaxy via ram pressure stripping \citep{gunn72}.
However, the relative importance of different AGN triggering mechanisms, including major and minor mergers and secular processes such as bar instabilities and galaxy interactions, remains an open question.

Conversely, the AGN can also impact the galaxy cluster environment. Radio-mode feedback plays an important role in regulating the cooling of hot gas from the ICM and limiting star formation. The feedback scenario is used to explain the lack of star formation in the inner cluster galaxies. In radio-mode, the energy from the AGN is deposited in the ICM by radio-emitting jets, creating low-density bubbles \citep[e.g.][]{peterson03,fabian12}. Observations of the X-ray cavities in clusters, formed by the transfer of mechanical energy into the ICM, suggest that AGN regulate cluster cooling at redshift $z\sim1$ during major assembly of the cluster and ICM \citep{hlavacek-larrondo15,mcdonald15}.

The literature on the frequency of AGN in galaxy clusters is mixed on how the environment impacts AGN triggering. The measured cluster AGN content can depend on factors such as the redshift, mass, and AGN selection method used in the study. Early work showed AGN are less likely to reside in galaxy cluster environments at $z\sim0.04$ \citep{dressler85}. However, more recent studies have shown the number of AGN in clusters and the ratio of AGN to cluster galaxies increases with higher redshift \citep{galametz09,martini09,martini13,pentericci13,bufanda17}. 

Early surveys also relied on optical spectroscopy, which is insensitive to obscured, Compton-thick AGN. With the addition of multi-frequency AGN surveys, a more nuanced picture of AGN in galaxy clusters emerges. Mid-infrared (MIR) selection has the benefit of being able to detect AGN obscured by dust. Ultraviolet light reprocessed by the dusty torus is bright in MIR bands but would otherwise be missed by optical selection \citep[e.g.,][]{lacy04, stern12, donley12}. Selection by radio and X-ray emission stemming from the AGN's nuclear activity is also advantageous in that sources that emit strongly in these wavelengths are very likely to be AGN. \citet{galametz09}, using AGN catalogs selected in the radio, infrared, and X-ray, report an increase in the amplitude of overdensity of AGN in clusters as a function of increasing redshift. \citet{croft07} estimate that $20\%$ of all low-redshift galaxy clusters have a radio-detected brightest cluster galaxy (BCG). Multiple studies using X-ray data show an excess of AGN in the central regions of galaxy clusters \citep[e.g.][]{ebeling01, ruderman05, ehlert13}.

Detailed observations reveal variations in the number of cluster AGN as a function of clustercentric distance. \citet{pimbblet13}, studying six galaxy clusters at $z\sim0.06$, find that the fraction of spectroscopically-confirmed optically-selected AGN increases from cluster center to $\sim2r_\mathrm{virial}$, beyond which the AGN fraction declines. In $42$ clusters at $0.2<z<0.7$, \citet{ehlert14} calculate an X-ray AGN fraction that is $\sim1.5-3$ times lower than the expected AGN fraction in the field in the innermost $\sim{r_{500}}$ and converges to expected field values beyond $\sim2r_{500}$.  \citet{haines12} find that the X-ray selected AGN population in a sample of 28 massive clusters at $0.15<z<0.30$ are dynamically associated with an infalling population and are preferentially aligned with the cluster caustics, and shows the efficacy of the cluster environment at suppressing nuclear activity in cluster members. Several authors also find tentative evidence for a secondary peak of AGN near the virial radius of the galaxy cluster \citep{ruderman05,fassbender12,koulouridis14}, which they attribute to an enhancement of the merger fraction in the cluster infall region. \citet{ehlert13} find weak evidence for a secondary excess between $1.5-2 r_{500}$ in all X-ray bands for the X-ray source radial profile in $43$ ROSAT All Sky Survey galaxy clusters, but did not find higher rates of AGN triggering near the virial radius as compared to the field when their sample was expanded to $135$ clusters \citep{ehlert15}. \citet{fassbender12} suggest that the secondary excess could be a function of cluster mass, where more massive galaxy clusters have a higher velocity dispersion that is less conducive to galaxy mergers.

We are motivated to revisit the study of AGN in galaxy clusters with a much larger cluster sample at high redshift. This paper presents a study of AGN at the positions of 2300 galaxy clusters at $z\sim1$ discovered with the Massive and Distant Clusters of WISE (MaDCoWS) survey. Section~\ref{sec:data} introduces the MaDCoWS data along with the AGN catalogs. We discuss the methods and results of the distribution of cluster AGN overdensities and fractions in Sections~\ref{sec:agndist} and \ref{sec:agnfrac}. In Section~\ref{sec:cluprop}, we investigate the dependence of AGN on cluster mass and central radio activity. We discuss the implications of our work in Section~\ref{sec:discussion}.

Throughout the paper, we adopt the nine-year Wilkinson Microwave Anisotropy Probe (WMAP9) cosmology of $\Omega_\mathrm{M}=0.287$, $\Omega_{\Lambda}=0.713$, and $H_0 = 69.32$~km~s$^{-1}$~Mpc$^{-1}$ \citep{hinshaw13}. We convert angular distances to physical distances assuming $z=1$. Unless otherwise stated, all magnitudes are in the Vega system. 


\section{Data}
\label{sec:data}

\begin{deluxetable}{lcccc}
\tabletypesize{\footnotesize}
\tablecolumns{5}
\tablecaption{Number of Galaxy Clusters within AGN Catalog Footprints\label{tab:clusters}}
\tablehead{\colhead{Cluster Sample}&\colhead{R09}&\colhead{A18} & \colhead{A18$\times$PS} & \colhead{FIRST}}
\startdata
MaDCoWS & 1063 & 1778 & 1703 & 1132 \\
~Spitzer Follow-up & 717 & 1217 & 1166 & 759 \\
~~$15\leq\lambda_{15}<22$ & 141 & 231 & 221 & 154 \\
~~$22\leq\lambda_{15}<40$ & 449 & 758 & 727 & 483 \\
~~$\lambda_{15}\geq40$ & 127 & 228 & 218 & 122 \\
Radio-Inactive & 747 & 920 & 920 & 920 \\
Radio-Active & 172 & 212 & 212 & 212 

\enddata
\end{deluxetable}

\subsection{Galaxy Cluster Sample}
\label{sec:clusample}

MaDCoWS \citep{gettings12,stanford14,brodwin15} identifies galaxy clusters as infrared-selected galaxy overdensities using the {\it Wide-field Infrared Survey Explorer} \citep[{\it WISE},][]{wright10}. Using the combination of {\it WISE} and Panoramic Survey Telescope and Rapid Response System \citep[Pan-STARRS,][]{kaiser02} data, MaDCoWS covers $17,668$~deg$^{2}$. At $z\sim1$, MaDCoWS provides the largest galaxy cluster sample at this epoch. For survey specifics, we refer the reader to \citet{gonzalez18}. In our analysis we use the $2300$ most significant MaDCoWS cluster candidates and refer to this sample as the ``MaDCoWS Cluster Catalog." For all clusters we use the WISE positions derived during the cluster detection process. 

This project also relies upon {\it Spitzer} IRAC 3.6 and 4.5~$\mu$m snapshot observations of $1956$ cluster candidates obtained during Cycles 9, 11, and 12 (PIs: Gonzalez, Brodwin; PIDs 90177, 11080, 12101).  We refer to these clusters as the ``{\it Spitzer} Follow-up Clusters." Along with overdensity confirmation, the {\it Spitzer} data enables photometric redshifts and richness determinations, which are described in \citet{gonzalez18}. Briefly, photometric redshifts are derived using the combination of $i-[3.6]$ and $[3.6]-[4.5]$ colors of galaxies within 1$^\prime$ of the cluster location. The redshift is determined via a comparison of the peak of the galaxy distribution in color-color space with the predicted colors for a passively evolving galaxy formed via a single stellar burst at $z_f=3$.  The richness $\lambda_{15}$ is then defined as the number of galaxies with $[4.5]>15 \mu$Jy ($\sim5\times10^{10} M_\odot$ at $z=1$)
in excess of the field density that lie within 1 Mpc of the cluster center and have $i-[3.6]$ and $[3.6]-[4.5]$ colors consistent with being possible cluster members. Specifically, the galaxy must be no more than one magnitude bluer in $i-[3.6]$ than the peak color of the galaxy distribution, and must be within $\pm0.15$ mag in $[3.6]-[4.5]$. The {\it Spitzer} follow-up clusters range in richness from $\lambda_{15}=0.1-122$, though the interquartile range is from $\lambda_{15}=22-36$ ($M_{500}=1.0-2.6\times10^{14}~M_{\odot}$).

\subsection{AGN Catalogs}

We use literature catalogs of AGN selected by various methods. The number of galaxy clusters considered for the analysis in each subset will vary due to the different coverage area of each AGN catalog. These values are listed listed in Table~\ref{tab:clusters}.

\begin{figure}
    \centering
    \includegraphics[width=\columnwidth]{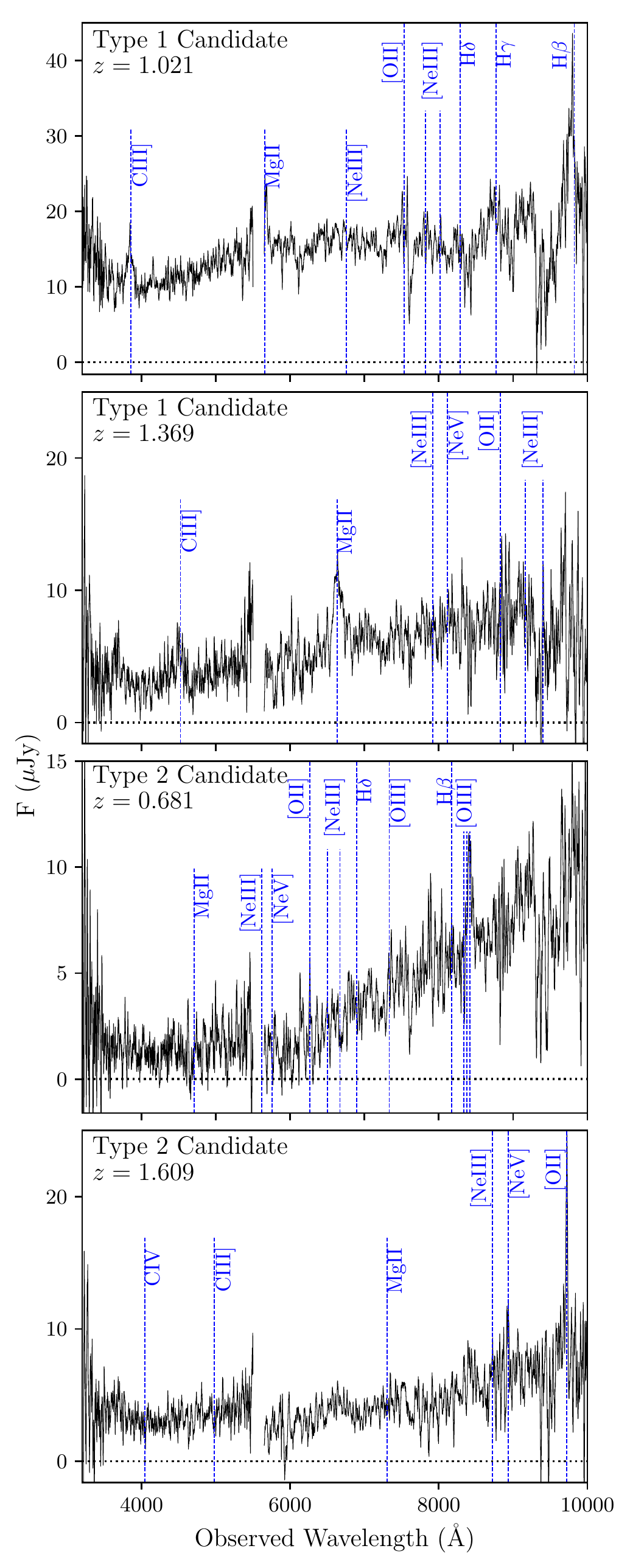}
    \caption{Spectra of two Type 1 and two Type 2 AGN candidates observed with Palomar DBSP on 22 November 2017. The Type 1 candidates were confirmed to be broad line AGN while the Type 2 AGN candidates are most likely narrow emission line galaxies.}
    \label{fig:spectra}
\end{figure}

\begin{figure}
    \centering
    \includegraphics[width=\columnwidth]{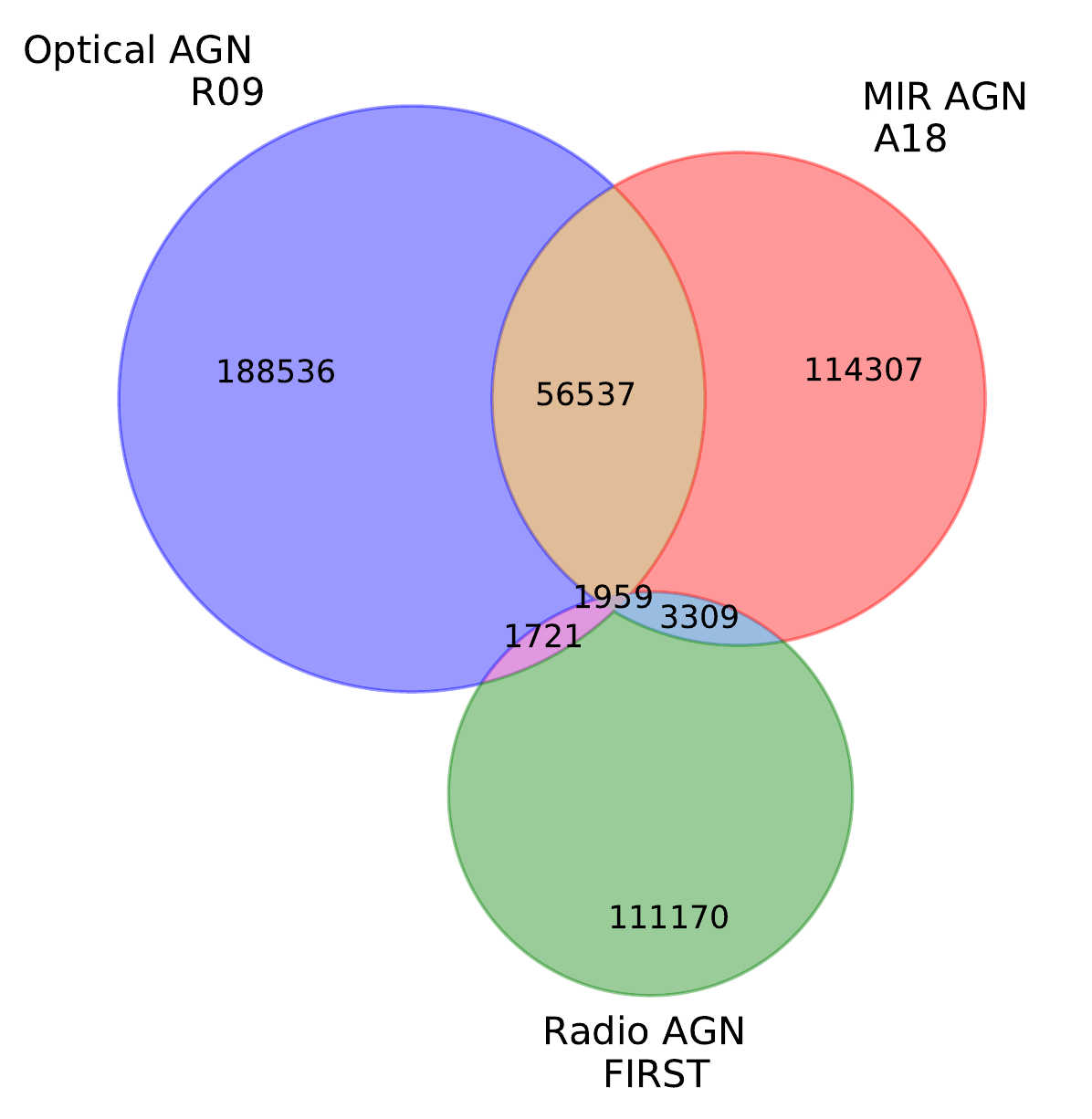}
    \caption{Overlap of sources in the AGN catalogs via optical-selection (R09), MIR-selection (A18) and radio signatures (FIRST) labeled by the number in each set.}
    \label{fig:catalogs}
\end{figure}

\subsubsection{Optical Selection}

\citet[][hereafter R09]{richards09} present a catalog of $1.2$ million quasars photometrically selected from the Sloan Digital Sky Survey (SDSS) Data Release 6. The catalog was constructed from SDSS point sources with extinction-corrected asinh magnitude $14.3<i<21.3$ (AB) from a 8417 deg$^{2}$ area. 
R09 employ the Bayesian algorithm presented in \citet{richards04} to classify objects as quasars or stars based on a training set of $75,382$ quasars from published catalogs. 

Using a method of auto-correlation of quasars presented in \citet{myers06}, R09 find a reliability of $71.5\pm3.5$\%. The completeness of the catalog to unobscured (Type 1) quasars is no worse than 70\%. A caveat is that the sample is relatively incomplete near $z\sim2.8$ and $z\sim3.5$, where the colors of stars and quasars are similar. It is also contaminated at $z\sim0.675$ by white dwarfs. R09 calculate an AGN catalog surface density of $141$~deg$^{-2}$. We refer to the R09 catalog as the ``optically-selected sample."

Though R09 also provides photometric redshifts for the catalog, accurate to ${\Delta}z=\pm0.3$ for 83\% of spectroscopically crossmatched sources, we choose not to select quasars based on redshift. In our analysis, we calculate the AGN field contribution by statistically subtracting foreground/background quasars. By not employing a photometric redshift cut, we also eliminate the possibility of selecting against quasars with miscalculated photometric redshifts.

\subsubsection{MIR Selection}

\citet[][hereafter A18]{assef18} have constructed a catalog using AllWISE data to photometrically select AGN via a two-color selection similar to that presented in \citet{assef13}. AllWISE is a combination of data from the {\it WISE} Full Cryogenic, 3-Band Cryogenic, and NEOWISE Post-Cryogenic survey. MIR selection distinguishes AGN from normal galaxies due to the infrared excess associated with AGN. In particular, MIR selection is less affected by dust extinction and can identify obscured, Compton-thick (Type 2) AGN missed by optical photometry. 

We use the A18 $90\%$ reliability catalog, where $90\%$ of catalog objects are expected to be bona fide AGN, containing more than $4.5$ million sources. To reduce the non-uniformity due to the varying depth of the WISE survey strategy, we only consider sources with $W2<15.5$. This limit ensures all catalog sources are brighter than the AllWISE $95$\% completeness in W2 magnitude of $15.7$ and excludes fainter elliptical galaxies that may have colors consistent with an AGN\footnote{\url{http://wise2.ipac.caltech.edu/docs/release/allwise/}}. We also avoid catalog sources in the range $-30^{\circ}\leq\delta\leq-10^{\circ}$, a region with higher density due to the South Atlantic Anomaly. We refer to the $2,010,062$ MIR-selected AGN in the A18 catalog matching our criteria as the ``MIR-selected sample." The surface density of the catalog is $86$~deg$^{-2}$.

The full A18 R90 catalog is optimized for 90\% reliability at the expense of completeness. Comparing the R90 catalog to the AGN and Galaxy Evolution Survey \citep[AGES,][]{kochanek12} in the NOAO Deep Wide-Field Survey (NDWFS) Bo\"otes field, the R90 selection recovers 17\% of the NDWFS Bo\"otes AGN. Because we only consider brighter sources in W2, our completeness should be higher than that of the full R90 catalog, though calculating the completeness and reliability at the W2 limit we consider is beyond the scope of this paper. However, with a similar selection function as A18, \citet{assef13} calculate a 53\% and 77\% completeness for their 90\% reliability catalog for sources brighter than $W2<15.73$ and $W2<15.05$, respectively.

\subsubsection{Optical-to-MIR Color Selection}

We further classify the MIR-selected AGN sample into Type 1 unobscured and Type 2 obscured AGN by employing the optical-to-MIR (OIR) color selection presented in \citet{hickox17} to distinguish MIR-selected AGN based on obscuration type. We crossmatch the MIR-selected sample to the Pan-STARRS DR1 catalog \citep{ps1} with a $2\arcsec$ matching radius to obtain $r$-band optical photometry, considering only sources with $5\sigma$ $r$ and W2 photometry. We choose to use PSF photometry from Pan-STARRS, as it extends to fainter magnitudes. For the MIR-selected sample, objects with colors bluer than $r-W2=3.1$ (AB) are deemed Type 1 and redder are Type 2.

From the MIR-selected sample of $2.0$ million AGN, $1.45$ million are within the Pan-STARRS detection region. We refer to this catalog as ``A18$\times$PS." $1.3$ million of these sources had a Pan-STARRS crossmatch within $2\arcsec$ of the WISE position. With our OIR color selection and $5\sigma$ $r$ photometry requirement, we find $805,335$ Type 1 AGN. $61$ had colors consistent with Type 1 AGN but lower than $5\sigma$ detection, and we exclude these AGN from our analysis. 

The remaining $646,945$ AGN are classified as Type 2. Of these, $344,382$ had Pan-STARRS photometry while $302,563$ did not have matches. We visually inspected a random selection of $50$ WISE AGN without Pan-STARRS counterparts. $45$ ($90\%$) of the sources appeared to be non-detections, where the WISE AGN was too faint to be detected by Pan-STARRS. The color selection requires Type 2 AGN to be fainter than $r>21.9$ (AB) while the $5\sigma$ limiting magnitude for stacked sources in Pan-STARRS is $r=23.2$ (AB). Thus, any MIR-selected AGN not detected by Pan-STARRS would have colors redder than $r-W2=3.1$ (AB). For the remaining five sources, three appeared to be multiple Pan-STARRS sources blended into one WISE source, due to the limitations of the WISE PSF, and two had bad WISE positioning or were in areas of poor Pan-STARRS data. From this analysis, we conclude that $10$\% of WISE AGN that did not match with a Pan-STARRS counterpart are mismatched due to blending or bad data. If we assume that an equal number of these are Type 1 and Type 2, then only $5$\% of Type 2 AGN are misidentified, and would not significantly alter our later results.

We also use a combination of publicly available SDSS I/II and BOSS spectra and targeted observations with Palomar Observatory to spectroscopically confirm a sample of photometrically-selected Type 1 and Type 2 AGN candidates as broad- or narrow-line AGN, respectively. We find that $90$\% of Type 1 candidates with SDSS I/II or BOSS spectra were classified as broad-line quasars.  We targeted two Type 1 candidates with the Palomar Double Spectrograph (DBSP) instrument. The spectra of both targets are consistent with being broad line sources.

Type 2 AGN, given their obscured optical emission, are fainter in optical bands than Type 1 AGN, and are less likely to be targeted for SDSS spectra. 
However, we obtained spectra of two Type 2 candidates with Palomar DBSP, both of which were confirmed to be narrow line objects. The Palomar DBSP spectra are shown in Figure~\ref{fig:spectra}.

\subsubsection{Radio Source Catalog}

We use the 2014 December 17 version of the Faint Images of the Sky at 20-cm \citep[FIRST,][]{becker95} Catalog of radio sources at $1.4$~GHz, collected with the Very Large Array (VLA). 
The catalog contains $9.46\times10^5$ sources over $10,575$ deg$^{2}$ with a $1$ mJy limiting flux density threshold and a $5\arcsec$ ($40.6$ kpc at $z=1$) resolution. 

To avoid low quality sources near edges of catalog coverage, we limit the catalog to regions $110^\circ<\alpha<262^\circ$ and $-8^\circ<\delta<57^\circ$ in the North and $325^\circ<\alpha<40^\circ$ and $-10^\circ<\delta<10^\circ$ in the South. We only consider sources with low sidelobe probability ($\mathrm{SIDEPROB}\leq0.015$) and with integrated flux above the limiting flux threshold ($\mathrm{FINT}\geq1$~mJy). We find $620,231$ FIRST catalog sources that match our criteria, and a source density of $62$~deg$^{-2}$.

The FIRST flux limit at $z=1$ corresponds to a $1.4$ GHz luminosity threshold of $L_{1.4~\mathrm{GHz}}=4\pi{S}D_{A}^2(1+z)^{3+\alpha}=4.7\times10^{24}$~W~Hz$^{-1}$, assuming spectral index\footnote{Defined $S\propto\nu^{-\alpha}$.} $\alpha=0.8$. Selecting radio sources with $L_{1.4~\mathrm{GHz}}\gtrsim10^{24}$~W~Hz$^{-1}$ provides high reliability of tracing AGN. \citet{mauch07} find that all $593$ radio sources with $L_{1.4~\mathrm{GHz}}>10^{24}$~W~Hz$^{-1}$ detected by the NRAO VLA Sky Survey and crossmatched with sources brighter than $K=12.6$ in the Second Incremental Data Release of the 6 degree Field Galaxy Survey (6dFGS DR2) were spectroscopically confirmed to be radio AGN. Catalog completeness is expected to be low. Canonically, only $\sim10$\% of all AGN exhibit radio emission \citep[e.g.,][]{white00}.

\subsubsection{AGN Identified in Multiple Catalogs}

Invariably some sources are selected as AGN in multiple catalogs. We consider the fraction of multiply-selected AGN in a $1841$~deg$^{2}$ region (RA: $150<\alpha<200$, Dec: $0<\delta<40$) with coverage by all three AGN catalogs. Figure~\ref{fig:catalogs} shows the overlap between the AGN catalogs crossmatched with a $1\arcsec$ radius.


\section{AGN Distribution}
\label{sec:agndist}

\begin{figure*}
    \centering
    \includegraphics[width=\textwidth]{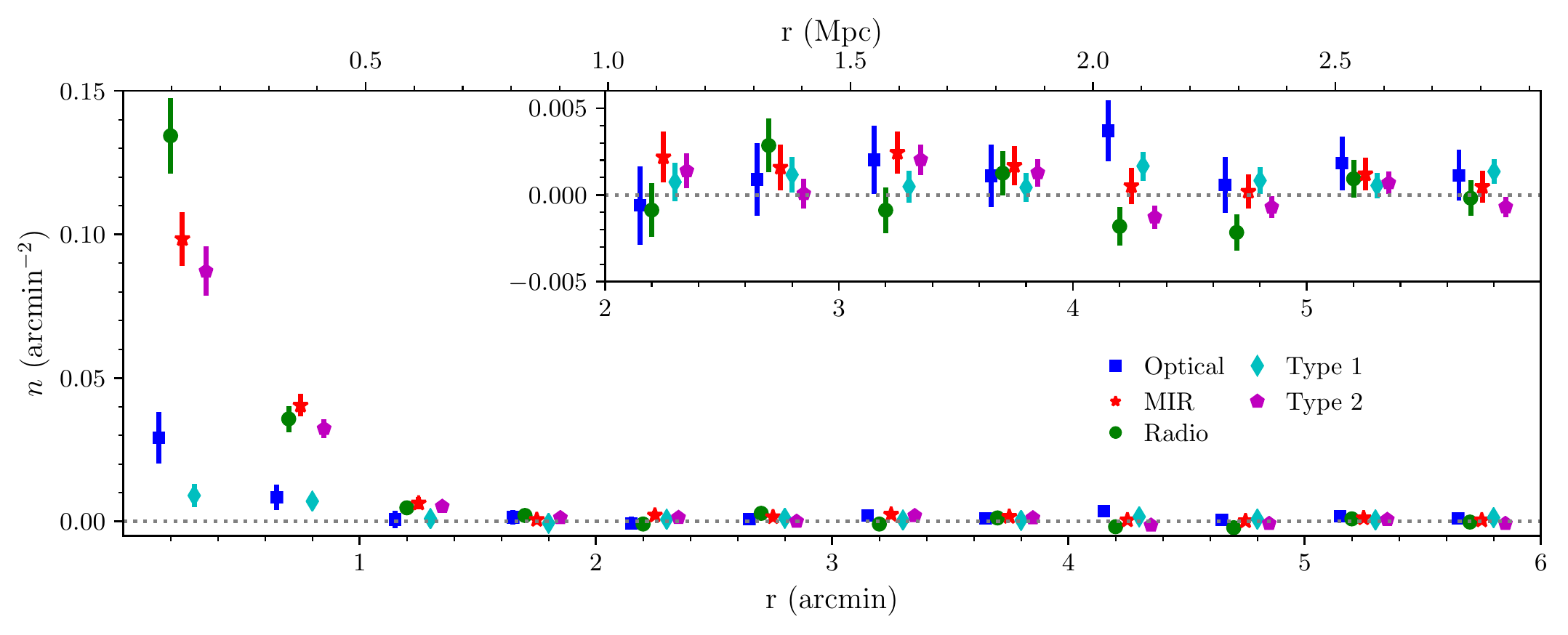}
    \caption{The cluster AGN excess as a function of clustercentric distance for the MaDCoWS cluster sample within $6\arcmin$ of cluster center. The x-axis is offset for clarity. We find an excess in AGN in the central $1.5\arcmin$ region. Beyond $2\arcmin$, the AGN excess returns to field levels. The inset shows the minimal variation outward of $r=2\arcmin$.}
    \label{fig:agndist}
\end{figure*}

\begin{figure*}
    \centering
    \includegraphics[width=.7\textwidth]{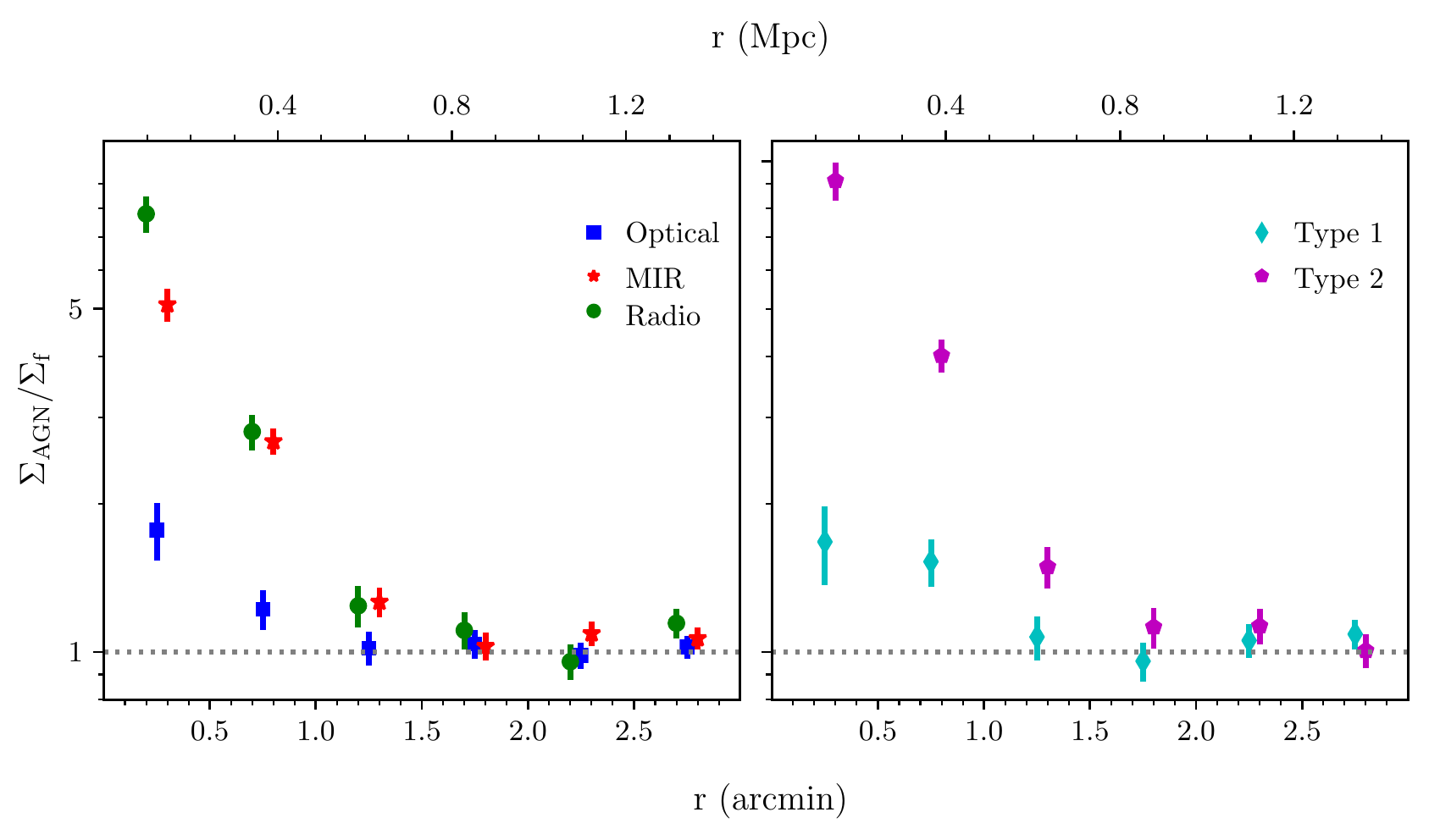}
    \caption{The projected AGN surface density distribution as a function of clustercentric distance. Accounting for varying densities amongst the AGN catalogs, the radio-selected AGN are most overdense in cluster environments within the central $1\arcmin$. Obscured AGN are also more likely to be found in clusters than unobscured AGN.}
    \label{fig:agndist_normed}
\end{figure*}

\begin{deluxetable}{cccc}
\tabletypesize{\footnotesize}
\tablecolumns{4}
\tablecaption{Field AGN Density \& Fraction\label{tab:agndensity}}
\tablehead{\colhead{AGN Catalog}&\colhead{$\eta_\mathrm{f}$}&\colhead{$\eta_\mathrm{f,SDWFS}$}& \colhead{$f_\mathrm{f}$ } \\
\colhead{} & \colhead{($10^{-2}$~arcmin$^{-2}$)} &\colhead{($10^{-2}$~arcmin$^{-2}$)} & \colhead{($10^{-3}$)} }
\startdata
R09 & $ 3.76 \pm 0.02$ & $3.29 \pm 0.11$ & $4.2 \pm 0.2$ \\
A18 & $ 2.47 \pm 0.01$ & $2.58 \pm 0.10$ & $3.4 \pm 0.2$ \\
FIRST & $1.72\pm0.01$ & $2.16\pm0.09$ & $1.4\pm0.1$ \\
Type 1 & $ 1.29 \pm 0.01 $ & $1.54 \pm 0.08$ & $1.8\pm0.1$ \\
Type 2 & $ 0.42 \pm 0.01 $ & $0.35 \pm 0.04$ & $0.6\pm0.1$ 
\enddata
\tablecomments{Column 2: The average density of the AGN within $5\arcmin-20\arcmin$ of annular region of cluster centers. Column 3: Density of AGN in SDWFS region. Column 4: The field AGN fraction after accounting for AGN catalog non-uniformity (described in Section~\ref{sec:agnfrac_method}).}
\end{deluxetable}

\subsection{Stacking Analysis}

We determine AGN radial distributions via a stacking analysis. First, we calculate the projected surface density of AGN within a radial annulus around cluster centers. 
For each radial bin of area $A(r)$, the projected AGN surface density is \begin{equation}
    \eta_\mathrm{AGN}(r)=\sum\limits_{i=1}^{N_\mathrm{cl}}N_\mathrm{AGN,i}(r)/A(r),
\end{equation}
where $N_\mathrm{AGN,i}(r)$ is the number of AGN found within the radial bin for cluster $i$ and $N_\mathrm{cl}$ is the total number of clusters. We then subtract a constant foreground/background AGN contribution, defined as $\eta_\mathrm{f}$, taken to be the projected AGN surface density within a radial bin of $5\arcmin-20\arcmin$. We consider any AGN surface density after subtracting the field contribution to be associated with the cluster. The cluster AGN excess, $n$, is then defined as
\begin{equation}
    n(r) = \frac{\eta_\mathrm{AGN}(r)-\eta_\mathrm{f}}{N_\mathrm{cl}}.
\end{equation}
Note that $N_\mathrm{cl}$ will vary based on the AGN selection, due to the different footprints of the various catalogs. $N_\mathrm{cl}$ for all AGN catalogs considered is listed in Table~\ref{tab:clusters}.

We estimate the uncertainty assuming the Poisson statistics for $N_\mathrm{AGN,i}(r)$ and $N_{f}$, defined as the total number of AGN within the radial bin and background annulus of $N_\mathrm{cl}$ clusters, respectively, and $N_\mathrm{cl}$. Explicitly, the uncertainties are equal to the square root of the quantities. We then estimate the uncertainty on $n$ for each radial bin by error propagation.
We have used the CHASC code from \citet{park06} to confirm that this approximation is valid even in bins with the lowest number of AGN.

\subsection{Radial Profiles}

In Figure~\ref{fig:agndist}, we plot the radial distribution of the excess AGN density per cluster as a function of cluster-centric distance. We note that the distribution of AGN excess in all AGN catalogs shows a positive signal within the central $r\lesssim1.5\arcmin$. 
The radio- and MIR-selected AGN signal is significantly higher in the cluster center than that of the optically-selected. The right panel, comparing the radial distributions for AGN selected by their OIR color, shows that the AGN excess for Type 1 and 2 AGN deviates in the central $1\arcmin$ region. The highest significance AGN excess for the distribution within $2\arcmin$ is $3.3\sigma$, $11\sigma$, and $10\sigma$ for optically-, MIR-, and radio-selected AGN, and $2.5\sigma$ and $4.2\sigma$ for Type 1 and 2, respectively, occurring in the central-most radial bin for all AGN. Though all AGN are overdense in the central cluster region, the AGN excess is still equivalent to less than $1$~AGN per cluster.

Beyond $r\sim2\arcmin$ ($1$~Mpc), the AGN density converges to the field value. The inset of Figure~\ref{fig:agndist} highlights the minimal variation in all AGN catalogs in the outer cluster region from $r=2\arcmin-6\arcmin$. We do not observe a secondary peak at the infall radius (between $4\arcmin-6\arcmin$ or $1.9-2.9$~Mpc). The highest absolute significance values in the outer region are $2.1\sigma$, $2.0\sigma$, $2.1\sigma$, $1.8\sigma$, and $2.3\sigma$ for optically, MIR, radio, Type 1, and Type 2 AGN, respectively, while the average within this radial range is between $0.8\sigma-1.3\sigma$.

Radio AGN are often complex and contain multiple components such as jets and cores. These components may be detected as multiple sources in the FIRST survey and double-counted by our analysis. We address this by visually inspecting galaxy clusters with multiple radio sources. We consider clusters with multiple radio-selected AGN within $1\arcmin$, the region containing the bulk of the cluster AGN population. Within the $59$ clusters with multiple radio-selected AGN, we find $130$ radio sources within the central $1\arcmin$. Upon visual inspection of the radio emission morphology for lobes and jets, we identify $24$ radio sources that are most likely a component of another radio-selected AGN. We conclude that only $11$\% of central $1\arcmin$ radio sources are double-counted and therefore do not significantly alter our overall results.

Blending can be an issue where the coarse resolution of WISE can confuse multiple adjacent galaxies for one galaxy. This is particularly true in the inner cluster regions for the A18 and A18$\times$PS catalogs. For our analysis, blending would decrease the number of AGN in the MIR-selected, Type 1, and Type 2 AGN excess especially in the central $r<0.5\arcmin$ cluster region. However, this effect would be the same between Type 1 and 2 AGN and is not responsible for the divergence in the core.

To normalize out the different source densities of the AGN catalogs, we also show the ratio of projected AGN surface density to the background field level in Figure~\ref{fig:agndist_normed}. The MIR- and radio-selected AGN surface densities are higher in cluster centers than the optically-selected AGN, by a factor of $5.1$ and $7.8$ relative to the field in the inner radial bin, respectively. The right panel compares the distribution of the field-relative projected surface density of Type 1 and Type 2 AGN in MaDCoWS clusters. Our results show that AGN with redder OIR colors are $2.5-5$ times more likely to be found within the central $1\arcmin$ region of cluster centers than AGN with bluer OIR colors.

\section{AGN Fraction}
\label{sec:agnfrac}

\begin{figure*}
    \centering
    \includegraphics[width=.7\textwidth]{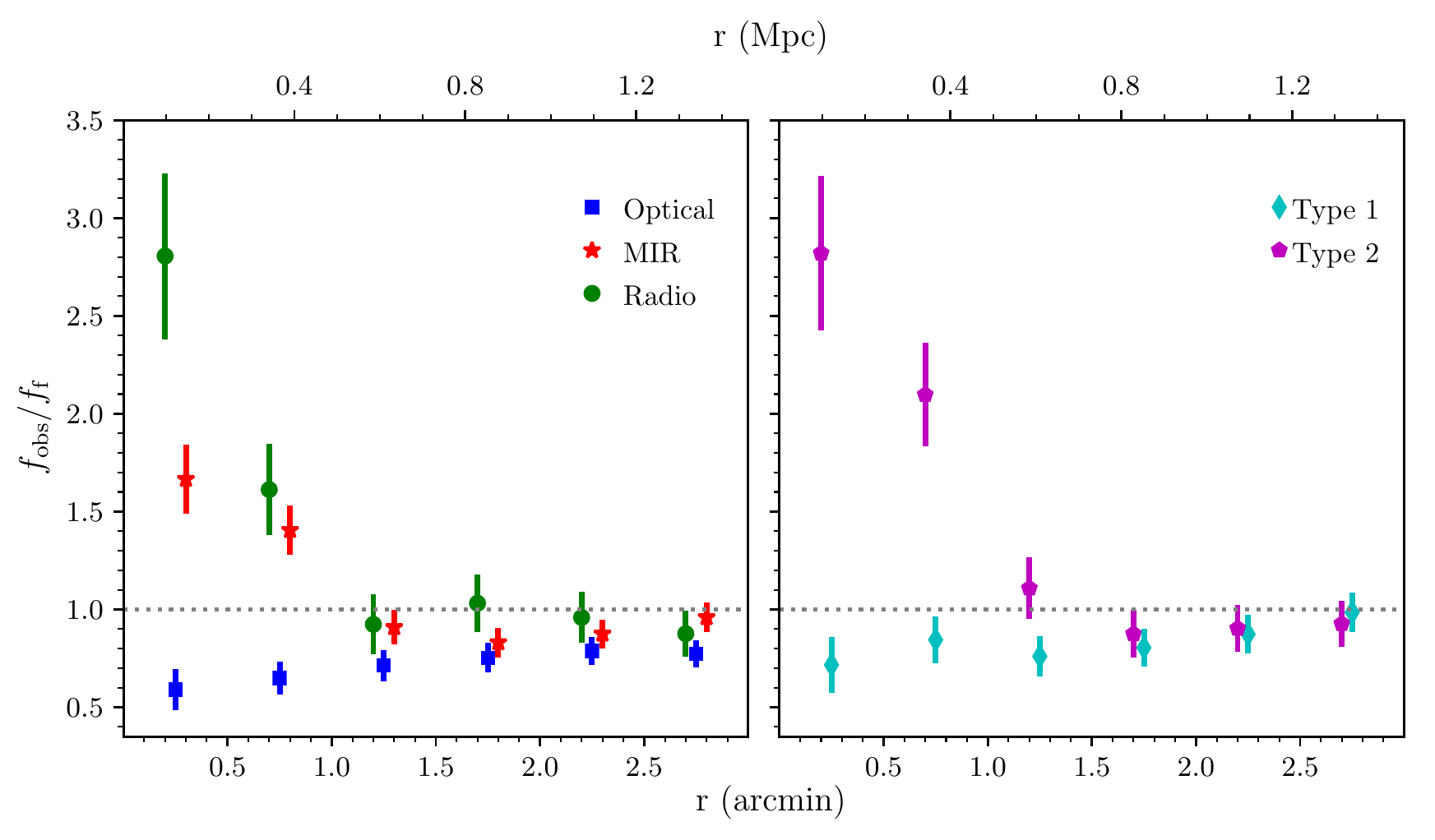}
    \caption{The observed AGN fraction divided by the field fraction as a function of clustercentric distance. Deviation from the field level are in the central $1\arcmin$ region and are dependent on the selection method of the AGN catalog.}
    \label{fig:agnfrac}
\end{figure*}

\begin{figure*}
    \centering
    \includegraphics[width=.7\textwidth]{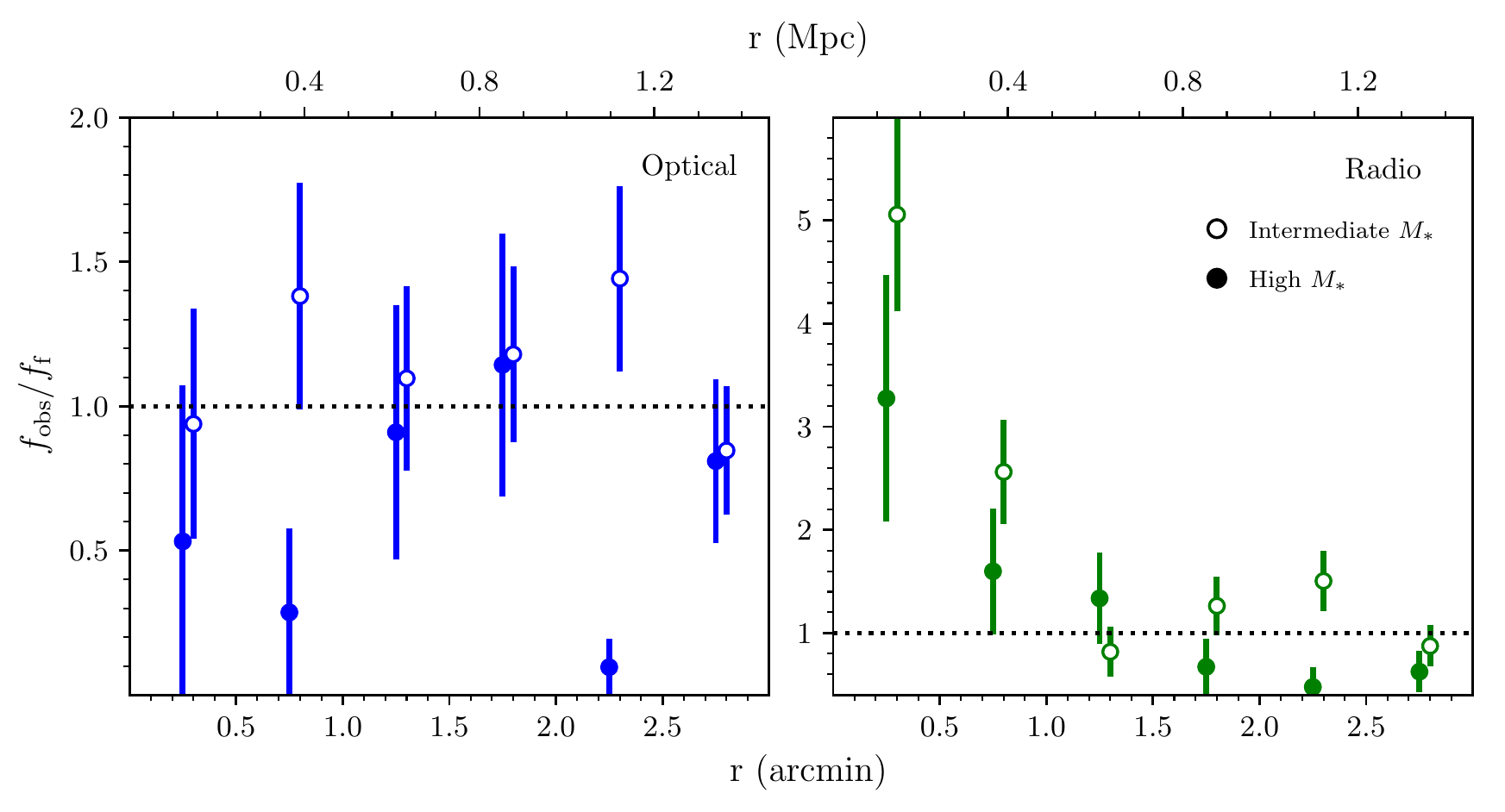}
    \caption{The field-relative observed AGN fraction for optically- and radio-selected AGN as a function of clustercentric distance when only considering galaxies with intermediate stellar mass ($3\times10^{10}~M_{\odot} <M_{*}\leq4\times10^{11}~M_{\odot}$, open circles) and high stellar mass ($M_{*}>4\times10^{11}~M_{\odot}$, closed circles). Cluster galaxies of intermediate stellar mass are more likely to host radio-selected AGN in the central $1\arcmin$ than galaxies of high stellar mass.}
    \label{fig:agnfracgalflux}
\end{figure*}

To decipher whether the rise in amplitude of excess AGN as a function of radius is due solely to the increasing cluster galaxy density towards cluster center or a true change in AGN frequency, we calculate the fraction of cluster galaxies that host an AGN as a function of clustercentric distance.

\subsection{Methodology}
\label{sec:agnfrac_method}
The cluster AGN fraction is defined simply as the ratio of cluster AGN to cluster galaxies. For {\it Spitzer} follow-up clusters, we identify candidate cluster members by selecting galaxies in IRAC $4.5\mu$m with minimum flux of $10\mu$Jy ($[4.5]<18.14$, equivalent to a stellar mass of $M_{*}\gtrsim3\times10^{10}$~$M_{\odot}$), and color $[3.6]-[4.5]>0.1$. This selection is designed to exclude field stars and minimize foreground galaxy contamination. We crossmatch selected galaxies to the R09, A18, FIRST, and A18$\times$PS catalogs. We use a $1\arcsec$ radius to match {\it Spitzer} catalogs to the R09, a $1.5\arcsec$ radius to match to the A18 and A18$\times$PS catalogs to account for the positional uncertainty in WISE, and a $2\arcsec$ crossmatching radius to account for radio sources with extended morphologies. We do not expect radio source host ambiguity to severely affect our results. After visual inspection of the radio sources found within the central $1\arcmin$ of clusters with {\it Spitzer} IRAC photometry, we only find $6$\% of radio sources  where we cannot identify the host galaxy. In these cases, the lack of counterpart identification is not due to the confusion between multiple sources, but rather that the host galaxy is fainter than the detection threshold.

We use a $7.25$~deg$^{2}$ region in the {\it Spitzer} Deep, Wide-Field Survey \citep[SDWFS,][]{ashby09} to obtain the field galaxy count. We limit the SDWFS field to $216.25^\circ<\alpha<218^\circ$ and $32.4^\circ<\delta<35.6^\circ$ and $218^\circ<\alpha<219.5^\circ$ and $33.5^\circ<\delta<35.6^\circ$ to avoid regions with edge effects. We apply the same flux and color criteria to the SDWFS field and calculate a field galaxy level of $\zeta_\mathrm{f,SDWFS}=7.174\pm0.017$~arcmin$^{-2}$.

Because the depth of each AGN catalog may vary across the sky, the number of AGN observed in the SDWFS field may not represent that calculated in positions surrounding the galaxy clusters, due to the difference in surface density in these regions. We account for the density discrepancy in order to obtain the surface density of AGN hosted by galaxies in the field, $\zeta_\mathrm{AGN,f}$. For each AGN catalog, we apply the same AGN crossmatching algorithm to the SDWFS field galaxies to calculate the observed crossmatched surface density of AGN, $\zeta_\mathrm{AGN,fobs}$. We then scale $\zeta_\mathrm{AGN,fobs}$ by the ratio of AGN surface densities in the background annuli and SDWFS region. Explicitly, 
\begin{equation}
    \zeta_\mathrm{AGN,f}=\zeta_\mathrm{AGN,fobs}\times\bigg(\frac{\eta_\mathrm{f}}{\eta_\mathrm{SDWFS}}\bigg)
\end{equation}
where $\eta_\mathrm{f}$ is the average surface density of the AGN catalog within $5\arcmin-20\arcmin$ of all cluster centers and $\eta_\mathrm{SDWFS}$ is the surface density of the AGN catalog in the SDWFS region. The field AGN fraction is then defined as
\begin{equation}
\label{equ:agnfraceff}
    f_\mathrm{f} = \frac{\zeta_\mathrm{AGN,f}}{\zeta_\mathrm{f,SDWFS}}.
\end{equation}
Table~\ref{tab:agndensity} lists the AGN surface density in the cluster and SDWFS regions and the field AGN fraction per AGN catalog after accounting for catalog non-uniformity.

We lack spectroscopic redshifts to determine if the AGN or galaxy within our cluster galaxy selection is a foreground/background interloper or indeed a part of the cluster. Thus, the observed version of the AGN fraction $f_\mathrm{obs}$ is the ratio of the sum of the AGN and any interloping galaxies that match our selection criteria within the line of sight of the radial bin. For a radial bin of area $A(r)$, we define the observed AGN fraction
\begin{equation}
    f_\mathrm{obs}(r) = \frac{\sum\limits_{i=1}^{N_\mathrm{cl}}{\mathcal{N}_\mathrm{AGN,i}}(r)}{\sum\limits_{i=1}^{N_\mathrm{cl}}{\mathcal{N}_\mathrm{i}}(r)}
\end{equation}
where $\mathcal{N}_\mathrm{i}(r)$ and is the number of galaxies matching the selection criteria in the $i$th cluster and $\mathcal{N}_\mathrm{AGN,i}(r)$ is the number of AGN crossmatched with $\mathcal{N}_\mathrm{i}(r)$ galaxies per radial bin. The true cluster AGN fraction is calculated after subtracting the field contribution,
\begin{equation}
\label{equ:agnfrac}
    f_\mathrm{true}(r) = \frac{\sum\limits_{i=1}^{N_\mathrm{cl}}{(\zeta_\mathrm{AGN,i}(r)-\zeta_\mathrm{AGN,f})}}{\sum\limits_{i=1}^{N_\mathrm{cl}}{(\zeta_\mathrm{i}(r)-\zeta_\mathrm{f})}},
\end{equation}
where $\zeta_\mathrm{AGN,i}(r)=\mathcal{N}_\mathrm{AGN,i}(r)/A(r)$ and $\zeta_\mathrm{i}(r)=\mathcal{N}_\mathrm{i}(r)/A(r)$. 

\subsection{Results}

We present our AGN fraction results as the observed AGN fraction $f_\mathrm{obs}$ divided by the field AGN fraction $f_\mathrm{f}$. Towards larger radii, beyond the influence of the cluster, $f_\mathrm{obs}$ should approach $f_\mathrm{f}$. The left panel of Figure~\ref{fig:agnfrac} shows the field-relative observed AGN fractions for the R09, A18, and FIRST AGN catalogs as a function of clustercentric distance. The only region where the AGN fraction deviates from the field level is within $r<1.0\arcmin$. Beyond $r=1\arcmin$, the AGN fraction for all selection methods converges to a value consistent with the field fraction. The enhancement of  MIR- and radio-selected AGN in cluster cores may reflect more frequent triggering. Alternatively, gradients in galaxy masses, morphologies, and fuel supplies within the cluster may contribute to the observed central excesses.

The profile also differs by selection. The radio- and MIR-selected AGN fractions rise towards cluster centers while the distribution of the optically-selected AGN fraction decreases towards the centers of galaxy clusters. This result indicates that MIR- and radio-selected AGN are found more frequently in galaxy clusters while optically-selected AGN are less likely to favor galaxy cluster environments.

The field-relative observed AGN fraction for Type 1 and 2 AGN is also shown in Figure~\ref{fig:agnfrac}. The fraction of Type 1 AGN is at a deficit compared to field levels in cluster centers and gradually increases towards field levels with increasing radii. In contrast, the Type 2 AGN fraction is enhanced when compared to the field fraction within the central $1\arcmin$ and returns to field fraction levels beyond $r\sim1\arcmin$.

Of note is the similarity in the distributions of the field-relative observed AGN fraction of optically-selected and Type 1 AGN and of MIR-selected and Type 2 AGN. This reiterates that optically-selected AGN are mostly Type 1 AGN, and therefore unobscured objects, and MIR-selected AGN are a majority population of Type 2, obscured AGN. This also shows that, at $z\sim1$, dust-obscured AGN are found more readily in cluster centers while unobscured AGN are less likely to be in cluster environments.

\subsection{Dependence on Galaxy Stellar Mass}

The mean mass of galaxies increases towards the centers of galaxy clusters. However, the host galaxies of AGN are also more massive than galaxies without AGN \citep[e.g.][]{best05,xue10}. To disentangle the effects of increasing galaxy mass from those of the cluster environment on AGN triggering, we calculate the field-relative observed AGN fraction when considering a fixed stellar mass range. 

We can estimate the stellar mass from the {\it Spitzer} [4.5]-band flux, assuming all galaxies at redshift $z=1$. However, AGN can contaminate the stellar mass estimation. Thus, we only consider galaxies below that of the MIR color threshold of AGN ($[3.6]-[4.5]<0.6$) for optically- and radio-selected AGN. We also assume a formation redshift $z_{f}=3$, metallicity $Z=0.03$, \citet{conroy09} stellar synthesis population (SSP) model, and \citet{chabrier03} initial mass function (IMF).\footnote{Computed with EzGal: \url{http://www.baryons.org/ezgal/model.php}} We then segregate galaxies into two stellar mass bins: galaxies of intermediate stellar mass, $3\times10^{10}~M_{\odot} <M_{*}\leq4\times10^{11}~M_{\odot}$ (corresponding to $F_{4.5}=10-120$~$\mu$Jy), and high stellar mass, $M_{*}>4\times10^{11}~M_{\odot}$ ($F_{4.5}>120$~$\mu$Jy). These limits were chosen to ensure a large enough sample per stellar mass bin. We crossmatch the R09 and FIRST AGN catalogs to the stellar mass-limited cluster galaxy sample and recalculate the observed AGN fraction.

Figure~\ref{fig:agnfracgalflux} compares the field-relative observed AGN fraction considering cluster galaxies of intermediate and high stellar mass.
Our results show that cluster galaxies of intermediate stellar mass are more likely to host radio-selected AGN than those of high stellar mass. This result implies that the difference in cluster radio-selected AGN fraction compared to the field AGN fraction is not simply due to the increase in average galaxy mass towards centers of clusters. We do not find a difference in the optically-selected AGN fractions in intermediate versus high stellar mass cluster galaxies.


\section{Trends
with Cluster Properties}
\label{sec:cluprop}

Studies have shown that AGN triggering is also dependent on the properties of the cluster. In this section, we demonstrate the dependence of the cluster AGN excess and fraction on cluster richness and central radio activity.

\subsection{Cluster Richness Dependence}

\begin{figure*}
    \centering
    \includegraphics[width=\textwidth]{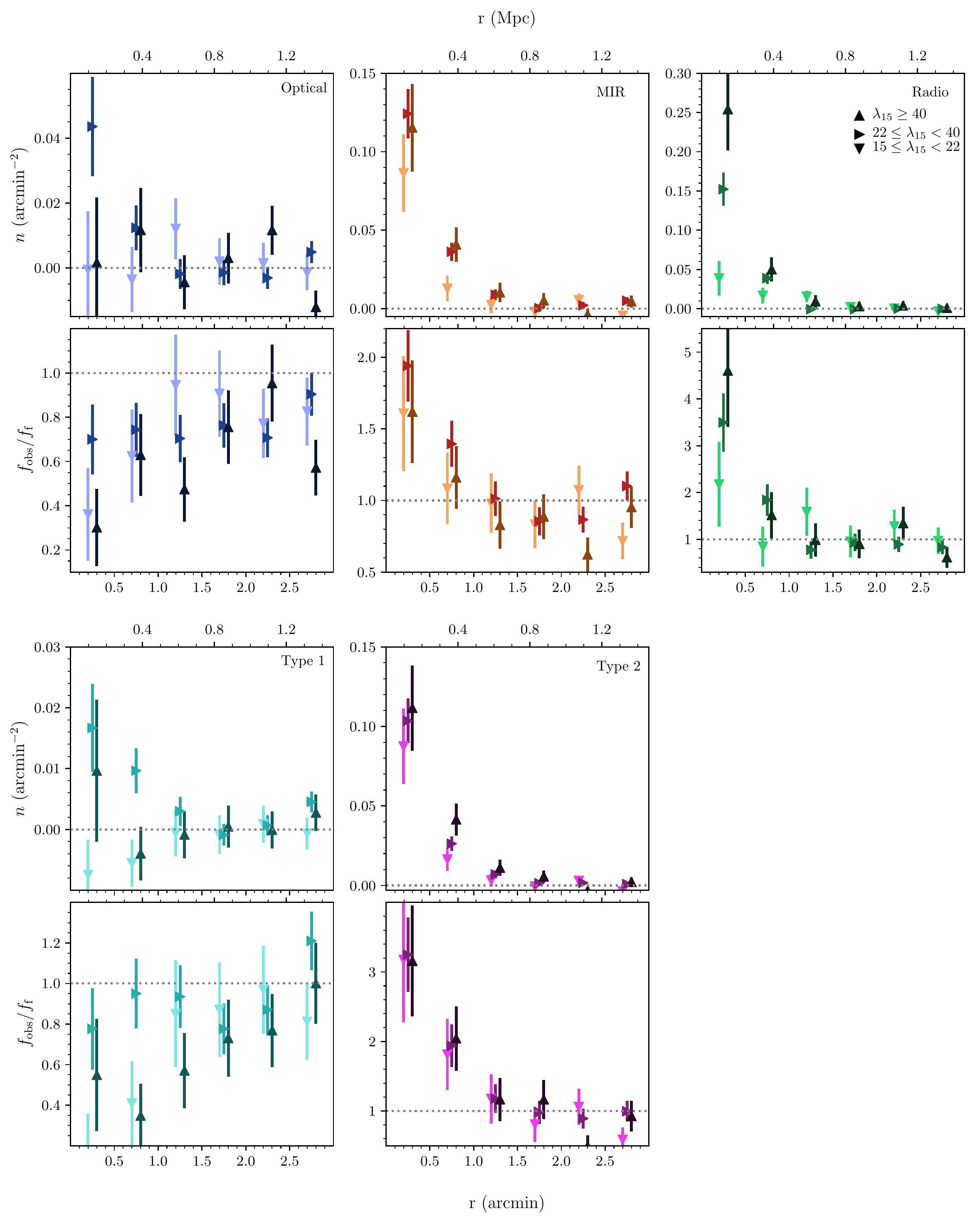}
    \caption{The radial distribution of the cluster AGN excess (first and third rows) and the observed AGN fraction relative to the field when dividing the sample of {\it Spitzer} follow-up galaxy clusters by cluster mass. We do not observe a trend with cluster richness.}
    \label{fig:agnplot_clumass}
\end{figure*}

\begin{figure}
    \centering
    \includegraphics[height=.905\textheight]{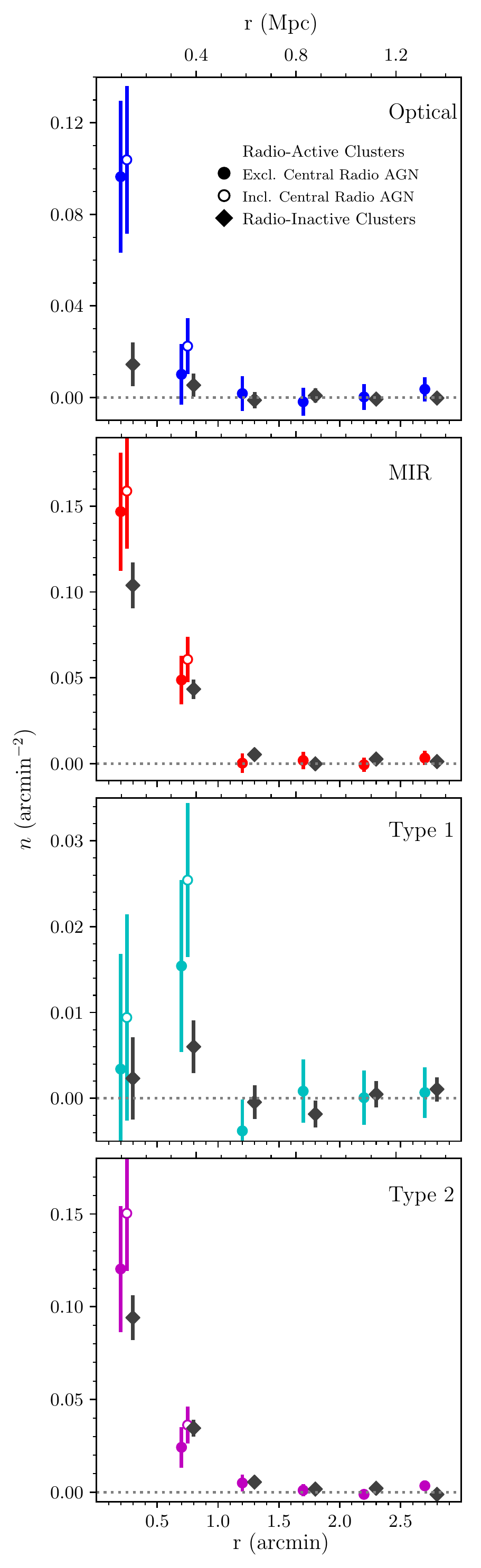}
    \caption{The comparison of the radial distribution of the cluster AGN excess in radio-active and radio-inactive galaxy clusters. Any AGN within the central $1\arcmin$ region that crossmatched with a FIRST source are subtracted from the excess (closed circles) to avoid selection bias. We find a higher cluster AGN excess in the central $1\arcmin$ of radio-active galaxy clusters than in radio-inactive clusters, most significant when considering optically-selected AGN.}
    \label{fig:agndist_rs}
\end{figure}

We divide clusters into low ($15\leq\lambda_{15}<22$, $9\times10^{13}~M_{\odot}{\leq}M_{500}<1\times10^{14}~M_{\odot}$), intermediate ($22\leq\lambda_{15}<40$, $1\times10^{14}~M_{\odot}\leq M_{500}<3\times10^{14}~M_{\odot}$) and high richness bins ($\lambda_{15}\geq40$)
and repeat the analysis presented in Sections~\ref{sec:agndist} and \ref{sec:agnfrac}.

The first and third rows of Figure~\ref{fig:agnplot_clumass} show the cluster AGN excess as a function of cluster richness, shown for each AGN population. Overall, the cluster AGN excess is above zero within $r<1.5\arcmin$ for all AGN selection methods and converges to field levels at $r>2\arcmin$. We do not find a strong dependence on cluster environment, as the AGN excess in clusters of each mass bin is statistically consistent at almost all locations and generally follows the same shape in profile. The only exceptions are for optically-selected and Type 1 AGN of intermediate richness clusters, for which there is modest evidence for an enhancement of AGN within the central $1\arcmin$ compared to the low- and high-richness clusters.

We also see no trend with cluster richness in the radial profile of the field-relative cluster AGN fraction as a function of cluster richness, shown in the second and fourth rows of Figure~\ref{fig:agnplot_clumass}. For each AGN population, the radial profile of the AGN fraction as a function of richness is similar in shape for all richness bins. Again, the exception is with optically-selected AGN in the centers of intermediate richness clusters, where the AGN fraction is flatter than the AGN fraction profile in low and high richness clusters.

\subsection{Correlations between Central Radio Activity and AGN Frequency}

We repeat the stacking analysis for clusters with and without central radio activity. MaDCoWS clusters with a radio source coincident within the central $1\arcmin$ are deemed ``radio-active clusters" while those without are ``radio-inactive." The total number of radio-active and radio-inactive clusters within the footprint of each AGN catalog is listed in Table~\ref{tab:clusters}. 

We note that, unlike in the calculations of the AGN fraction, we do not require a {\it Spitzer} galaxy association to denote that a cluster is radio active. Because the FIRST catalog does not provide associated redshifts, there is always the possibility that the radio source is not associated with the cluster but rather is a chance superposition. We calculate the probability for an intervening radio source to be part of our radio-active sample. 
If we offset the position of MaDCoWS clusters in any direction, we find that $5$\% of clusters have a radio source coincident with the central $1\arcmin$, equivalent to $55$ clusters in the FIRST footprint. Thus, we should expect that $26$\% of the galaxy clusters identified as radio-active to in fact be non-active.

We investigate whether the presence of a central radio source correlates with the number of AGN in the cluster center. Since we are selecting galaxy clusters known to contain radio AGN in cluster centers, we have an inherent bias if central radio-selected AGN are included in our analysis. Thus, we matched the R09, A18, Type 1 and 2 AGN catalogs with FIRST using a $2\arcsec$ matching radius and present both the AGN excess including and excluding the matched FIRST AGN within the central $1\arcmin$.

Figure~\ref{fig:agndist_rs} compares the excess of optically- and MIR-selected AGN and Type 1 and 2 AGN candidates in radio-active and radio-inactive clusters. All distributions for the radio-active clusters show excess AGN within the central $1\arcmin$ compared to that in radio-inactive clusters. The highest significance difference is in the optically-selected AGN where the number density of excess optical AGN in radio-active clusters in the innermost bin, excluding the crossmatched radio sources, is $0.096\pm0.033$~arcmin$^{-2}$ compared to $0.014\pm0.010$~arcmin$^{-2}$ in radio-inactive clusters. The excess of MIR-selected AGN and Type 1 and 2 AGN candidates also hint at a possible enhancement of the number of AGN in radio-active clusters. However, the large uncertainties in our calculations limit us from interpreting this result any further. 

We also compare the distributions of cluster richness of radio-active and radio-inactive clusters. The mean richness for radio-active and radio-inactive clusters is $\lambda_{15}=32.6\pm1.1$ and $\lambda_{15}=28.8\pm0.4$, respectively, implying that clusters with central radio activity are on average slightly more massive than radio-inactive clusters. A 2-sample Kolmogorov-Smirnov (KS) test calculates a p-value of $8.4\times10^{-4}$, indicating that the distributions of radio-active and radio-inactive clusters are formally inconsistent with being drawn from the same population. 
We repeat this analysis considering clusters with richness $15\leq\lambda_{15}<30$ and $\lambda_{15}\geq30$. Clusters of both richness bins show a higher amplitude of optically-selected AGN excess in radio-active clusters than in radio-inactive clusters in the central $1\arcmin$ region. Thus, we conclude that the optically-selected AGN excess in radio-active clusters is not simply an effect of the cluster richness.

We also examine if the AGN excess in radio-active vs radio-inactive galaxy clusters is related to the power of the central radio source. We split the radio-active cluster sample by total radio AGN $1.4$~GHz flux within the central $1\arcmin$ cluster region. However, we do not find a dependence of the amplitude of the cluster AGN excess on radio power in central cluster regions.

\section{Discussion}
\label{sec:discussion}

\begin{deluxetable*}{cccccc}
\tabletypesize{\footnotesize}
\tablecolumns{6}
\tablecaption{Cluster AGN Excess in Literature\label{tab:excess}}
\tablehead{\colhead{X-ray}&\colhead{$N_\mathrm{cl}$}&\colhead{Aperture}&\colhead{$z$}&\colhead{$N_\mathrm{AGN}$}&\colhead{Flux Limit} \\ & & & & & \colhead{($10^{-15}$~erg~cm$^{-2}$~s$^{-1}$)}}
\startdata
\citet{ehlert13} & $43$ & $r_{500}$\footnote{$1-1.7$~Mpc} & $0.2-0.4$ & $1.1\pm0.6$ & $5$ ($0.5-8$~keV) \\
\citet{gilmour09} & $148$ & $1$~Mpc & $0.1-0.9$ & $0.78\pm0.18$\footnote{\citet{gilmour09} also report an excess of $1.46\pm0.32$ without a given X-ray flux limit.} & $10$ ($0.5-8$~keV)\\
\citet{galametz09} & $140$ & $2\arcmin$ & $0.5-1.0$ & $0.09\pm0.11$ & $\sim7.8$ ($0.5-7$~keV)\\
\citet{galametz09} & $79$ & $2\arcmin$ & $1.0-1.5$ & $0.02\pm0.13$ & $\sim7.8$ ($0.5-7$~keV)\\
\hline\\
Radio & $N_\mathrm{cl}$ & Aperture & $z$ & $N_\mathrm{AGN}$ & Luminosity Limit \\
& & & & & ($10^{24}$~W~Hz$^{-1}$)\\
\hline
\citet{hart09} & $11$ & $1$~Mpc & $0.2-0.4$ & $\sim1.4$ & $0.3$ \\
\citet{gralla11} & $289$ & $0.5$~Mpc & $0.65-0.95$ & $0.10\pm0.02$ & $4.1$ \\
\citet{galametz09} & $121$ & $2\arcmin$ & $0.5-1.0$ & $0.09\pm0.04$ & $2.5$ \\
\citet{galametz09} & $69$ & $2\arcmin$ & $1.0-1.5$ & $0.23\pm0.09$ & $2.5$ \\
This Work & $1132$ & $1\arcmin$ & $\sim1$ & $0.19\pm0.02$ & $4.6$\\
This Work & $1132$ & $2\arcmin$ & $\sim1$ & $0.22\pm0.02$ & $4.6$\\
\hline\\
MIR & $N_\mathrm{cl}$ & Aperture & $z$ & $N_\mathrm{AGN}$ & Flux Limit\\
& & & & & ($\mu$Jy)\\
\hline
\citet{galametz09} & $140$ & $2\arcmin$ & $0.5-1.0$ & $0.02\pm0.04$ & $51$ ($5.8$~$\mu$m) \\
\citet{galametz09} & $79$ & $2\arcmin$ & $1.0-1.5$ & $0.09\pm0.06$ & $51$ ($5.8$~$\mu$m) \\
This Work & $2275$ & $1\arcmin$ & $\sim1$ & $0.17\pm0.01$ & $108$ ($4.6$~$\mu$m)\footnote{Considering the $W2<15.5$ magnitude limit applied to the A18 catalog.}\\
This Work & $2275$ & $2\arcmin$ & $\sim1$ & $0.20\pm0.02$ & $108$ ($4.6$~$\mu$m)\\
\hline\\
Optical & $N_\mathrm{cl}$ & Aperture & $z$ & $N_\mathrm{AGN}$ & Magnitude $i$ Limit \\
& & & & & (asinh)\\
\hline
This Work & $1063$ & $1\arcmin$ & $\sim1$ & $0.04\pm0.01$ & $14.5-21.3$\\ 
This Work & $1063$ & $2\arcmin$ & $\sim1$ & $0.05\pm0.02$ & $14.5-21.3$
\tablecomments{The excess number of AGN per cluster (Column 5) reported within the given aperture (Column 3).}
\end{deluxetable*}

\begin{figure}
    \centering
    \includegraphics[width=\columnwidth]{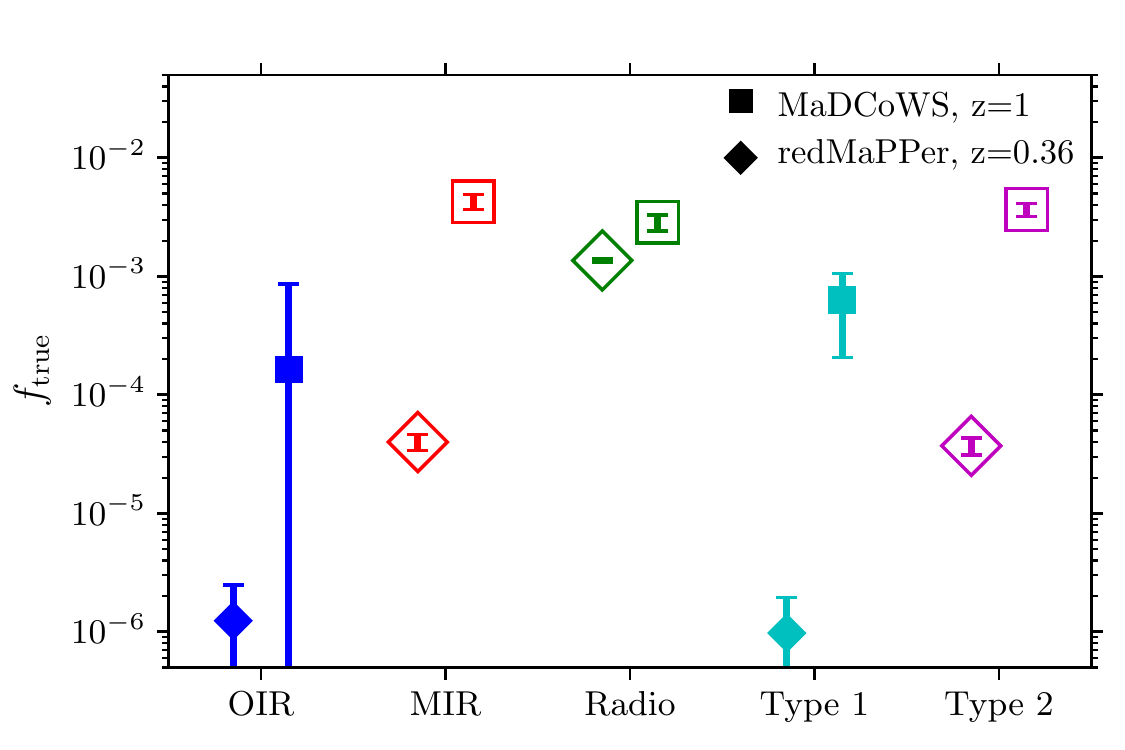}
    \caption{The true AGN fraction within a fixed aperture of $1$~Mpc for redMaPPer galaxy clusters with mean $z=0.36$ and MaDCoWS clusters at $z\sim1$. The selection criteria for AGN has been matched for the two epochs. The AGN fraction increases for all AGN types between $z\sim1$ and $z\sim0.36$.}
    \label{fig:agnfracevolution}
\end{figure}

\subsection{Comparison with Literature}

We have established that AGN density rises toward the centers of MaDCoWS galaxy clusters. Many authors have reported an increase in the projected AGN density at various wavelengths in the central regions of galaxy clusters across a range of redshifts \citep[e.g.,][]{galametz09,hart11,fassbender12,ehlert13}. We find that almost all the AGN density enhancement, regardless of selection method, is within $r=1-1.5\arcmin$ ($0.5-0.75$~Mpc) of the cluster center, which is in agreement with similar studies \citep[e.g.,][]{martini07,galametz09} though narrower than the distributions found with X-ray-selected sources \citep[e.g.,][]{ruderman05,gilmour09,ehlert13}. Table~\ref{tab:excess} presents a non-exhaustive list of cluster excess AGN measurements from the literature for a range of cluster redshifts and various AGN selection methods. We compare these studies to our results within $1\arcmin$ ($\sim0.5$~Mpc) and $2\arcmin$ ($\sim1$~Mpc), but we note that the AGN excess per cluster strongly depends on AGN selection type, luminosity, redshift and cluster mass considered in each study.

\citet{ruderman05} and \citet{fassbender12}, who studied X-ray sources in $51$ clusters at $0.3<z<0.7$ and $22$ clusters at $0.9<z<1.6$, respectively, find evidence for a secondary peak of AGN overdensity near $2-3$~$r_{200}$. However, this result was not observed in studies conducted with larger samples of galaxy clusters \citep[e.g.][]{gilmour09,ehlert14}. We also find no evidence for a secondary peak near the cluster infall region. \citet{koulouridis14} suggested that the secondary peak could depend on the cluster richness. However, our radial profiles when binned by cluster richness remain similarly flat. 

For optically-selected and Type 1 AGN, we find that the AGN fraction decreases within the central $1\arcmin$ ($0.5$~Mpc) cluster environment. The suppression is intensified closer to cluster centers. Both \citet{ehlert14} and \citet{pimbblet13} report suppression near cluster centers for X-ray and optically-selected AGN, respectively, but in lower redshift clusters. Our results confirm their findings, but with higher statistical precision and at a higher redshift regime. One possible explanation would be the cluster environment suppressing the triggering of AGN within the central $1\arcmin$ ($0.5$~Mpc) region.

We find that the enhancement or suppression of the cluster AGN fraction is very dependent on the AGN selection method. AGN selected by methods to identify more obscured type AGN, such as by the MIR or red OIR color, or by their radio signatures are more likely to be found in cluster environments. This result is in line with clustering studies that show Type 2 AGN are found in higher mass halos than Type 1 AGN \citep{hickox11,donoso14,dipompeo14,alevato14} and further adds to the evidence against the argument of obscuration purely due to orientation effects.

\subsection{Evolution of the AGN Fraction}

\citet{martini13} presented evidence that cluster AGN evolve more rapidly than field AGN, with cluster AGN fractions for luminous X-ray AGN increasing by a factor of $10$ from $z\sim0$ to $z\sim1$ and by another order of magnitude from $z\sim1$ to $z\sim1.5$. As a coarse test, we compare AGN fractions for galaxy clusters in the redMaPPer catalog \citep{rykoff14} at lower redshift to our AGN fractions.

The redMaPPer survey exploits the red sequence cluster-finding algorithm to detect galaxy clusters. We use version $6.3$ of the redMaPPer DR8 cluster and member catalog which applied the redMaPPer algorithm to the $\sim10,000$ deg$^{2}$ SDSS DR8 data. The catalog contains $\sim25,000$ galaxy clusters over redshift space $0.08<z<0.55$ with a completeness of $\gtrsim99\%$ at richness $\lambda_\mathrm{redMaPPer}>30$ and purity of $95\%$. 

We attempt to best replicate our MaDCoWS analysis with the redMaPPer clusters. To match the IRAC $4.5$ $\mu$m flux limit imposed on the MaDCoWS catalog, we also impose a similar magnitude limit on the redMaPPer member catalog. Given each individual redMaPPer cluster redshift, we convert the $10~\mu$Jy limit to a stellar mass of a passively evolving galaxy (assuming $z_{f}=3.0$, \citet{conroy09} SSP, and \citet{chabrier03} IMF) and calculate its corresponding SDSS $i$ magnitude. We only consider cluster members brighter in SDSS $i$ for each redMaPPer cluster. At $z=1$, we are only sensitive to radio sources with luminosity above $\sim5\times10^{24}$~W~Hz$^{-1}$ ($1$~mJy limiting flux in FIRST). Thus, we only consider FIRST sources above this luminosity limit at each redMaPPer cluster redshift. We also match the magnitude limits imposed upon the R09 ($i<21.1$ (AB)) and A18 catalogs ($W2<15.5$ (Vega)) for $z=1$ and $z=0.36$. We then crossmatched the redMaPPer cluster galaxy members to the AGN catalogs as described in Section~\ref{sec:agnfrac_method}. 

For redMaPPer clusters, we calculate AGN fractions within a fixed aperture of $1$~Mpc ($3.3\arcmin$ at $z=0.36$). Because the redMaPPer algorithm selects for red sequence members only, the field contribution to the AGN fraction is low, and the redMaPPer AGN fractions are more reflective of the true cluster AGN fraction rather than the observed AGN fraction for MaDCoWS. Thus, we compare redMaPPer AGN fraction values to the MaDCoWS true AGN fractions calculated with Equation~\ref{equ:agnfrac} for a radius $r=2\arcmin$ ($0.96$~Mpc at $z=1$). A comparison of the AGN fractions is shown in Figure~\ref{fig:agnfracevolution}.

Our results show that the cluster AGN fraction increases from $z=0.36$ to $z=1$. For the luminosity and magnitude limits considered, we find that the AGN fraction increases by at least a factor of $100$ for optical, MIR, Type 1 and Type 2 AGN and by a factor of $2$ for radio-selected AGN.
\citet{martini13} and \citet{bufanda17}, both studying X-ray AGN with $L_{X}>10^{43}$~ergs~s$^{-1}$, find that the AGN fraction increases by a factor of $7$ and $8$, respectively, between $0.1<z<1.0$.

\section{Conclusions}
\label{sec:conclusions}

We conducted a study of the AGN content in massive galaxy clusters at $z\sim1$. Employing the the largest known sample of galaxy clusters at $z = 1$ and catalogs of AGN selected by five distinct signatures, our results provide statistical power and nuance to previous work conducted with smaller samples. Our main results are as follows: 
\begin{itemize}
    \item We observe a rising density of AGN within the central $r<1.5\arcmin$ ($0.7$~Mpc) of MaDCoWS clusters. The AGN overdensity converges to field levels at $r=2\arcmin$ ($1.0$~Mpc). We find no evidence for a secondary peak of AGN overdensity beyond $r\gtrsim2\arcmin$.
    \item Optically-selected and Type 1 AGN decrease in AGN fraction towards cluster centers when compared to field AGN fraction levels while MIR-selected and Type 2 AGN show an enhancement within $r<1.5\arcmin$ compared to the field. Our results imply that cluster environments statistically contain more obscured than unobscured AGN, highlighting the importance of AGN selection when studying cluster AGN content.
    \item Radio-selected AGN are preferentially found in the inner $r<1\arcmin$ of cluster cores, regardless of the galaxy mass considered. This trend could be environmental in origin, but may also be a result of the morphology-density relation given that radio-selected AGN are preferentially found in massive elliptical galaxies \citep[e.g.,][]{best05}, or triggered by galactic mergers and interactions in the case of the central BCG \citep{chiaberge15}. 
    However, we also find that the AGN fraction is higher in the central $1\arcmin$ ($0.5$ Mpc) in host galaxies with stellar masses near $L_{*}$ compared with high mass hosts.
    \item Galaxy clusters with a radio source within $1\arcmin$ of cluster center show an enhanced number of optically-selected AGN as compared to clusters without central radio activity.
    Though radio-active clusters are on average more massive than radio-inactive clusters, we find that the excess of optically-selected AGN in radio-active clusters is still present even when matching the richnesses of the radio-active and radio-inactive cluster samples.
\end{itemize}


\section*{Acknowledgments}

The authors thank Ezequiel Treister and Roberto Assef for helpful discussion and the anonymous referee and statistical editor for feedback that improved the quality of this paper. Funding for this program is provided by NASA through the NASA Astrophysical Data Analysis Program, award NNX12AE15G, and the National Science Foundation grant AST-1715181. 
Parts of this work have also been supported through NASA grants associated with the {\it Spitzer} observations (PID 90177 and PID 11080).
W.M. was supported in part by the National Science Foundation Graduate Research Fellowship under Grant No. DGE-1315138. 

This publication makes use of data products from the Wide-field Infrared Survey Explorer, which is a joint project of the University of California, Los Angeles, and the Jet Propulsion Laboratory/California Institute of Technology, funded by the National Aeronautics and Space Administration. 
This work is also based in part on observations made with the Spitzer Space Telescope, which is operated by the Jet Propulsion Laboratory, California Institute of Technology under a contract with NASA. 
This research makes use of the Pan-STARRS1 Surveys (PS1) which have been made possible through contributions of the Institute for Astronomy, the University of Hawaii, the Pan-STARRS Project Office, the Max-Planck Society and its participating institutes, the Max Planck Institute for Astronomy, Heidelberg and the Max Planck Institute for Extraterrestrial Physics, Garching, The Johns Hopkins University, Durham University, the University of Edinburgh, Queen's University Belfast, the Harvard-Smithsonian Center for Astrophysics, the Las Cumbres Observatory Global Telescope Network Incorporated, the National Central University of Taiwan, the Space Telescope Science Institute, the National Aeronautics and Space Administration under Grant No. NNX08AR22G issued through the Planetary Science Division of the NASA Science Mission Directorate, the National Science Foundation under Grant No. AST-1238877, the University of Maryland, and Eotvos Lorand University (ELTE). 
This research made use of Astropy, a community-developed core Python package for Astronomy \citet{astropy}.


\bibliographystyle{apj}

\end{document}